\documentclass[useAMS,usenatbib]{mn2e}
\pdfoutput=1 
\usepackage{graphicx}
\usepackage{natbib}
\usepackage{amsmath}
\include{aas_journals}

\newcommand{\WNF}{\text{Panphasia}}
\newcommand{\MRG}{{\sc mrgk5-93}}
\newcommand{\ICgen}{{\sc ic\_2lpt\_gen}}

\newcommand{\RNG}{RAND}

\newcommand{\PRNG}{pseudorandom number generator}

\newenvironment{proto}{\begin{center} \large \tt \begin{minipage}{60em} }{\end{minipage}\end{center}}

\newcommand{\Realtype}{{\cal R}}
\newcommand{\Integertype}{{\cal I}}
\newcommand{\Statetype}{{\cal S}}
\newcommand{\Offsettype}{{\cal O}}

\newcommand{\Real}[1]{ {\tt REAL} \ {#1} }
\newcommand{\Integer}[1]{ {\tt INTEGER} \ {#1} }
\newcommand{\Offset }[1]{ {\tt RAND\_OFFSET}\  {#1} }
\newcommand{\State }[1]{ {\tt RAND\_STATE}\  {#1} }
\newcommand{\N}{{\tt Nstate}}

\newcommand{\Update}{{\tt Update}}
\newcommand{\Boost}{{\tt Boost}}
\newcommand{\Step}{{\tt Step}}
\newcommand{\Output}{{\tt Output}}
\newcommand{\Save}{{\tt Save}}
\newcommand{\Load}{{\tt Load}}
\newcommand{\Seed}{{\tt Seed}}

\newcommand{\Add}{+}
\newcommand{\Mul}{\times}

\title[] 
{Panphasia: a user guide.}

\begin{document}

\author[A. Jenkins \& S. Booth]{Adrian Jenkins$^1$\thanks{A.R.Jenkins@durham.ac.uk} \&
Stephen Booth$^2$ \\
$^1$Institute for Computational Cosmology, Department of Physics, University of Durham, 
South Road, Durham, DH1 3LE, UK\\
$^2$Edinburgh Parallel Computing Centre, The University of Edinburgh, JCMB, Edinburgh,
Mayfield Road, Edinburgh, EH9 3JZ, UK}
\maketitle

\begin{abstract}
 We make a very large realisation of a Gaussian white noise field,
called {\sc panphasia}, public by releasing software that computes
this field.  \WNF\ is designed specifically for setting up Gaussian
initial conditions for cosmological simulations and resimulations of
structure formation. We make available both software to compute the
field itself and codes to illustrate applications including a modified
version of a public serial initial conditions generator.  We document
the software and present the results of a few basic tests of the
field.  The properties and method of construction of \WNF\ are
described in full in a companion paper -- Jenkins 2013.
\end{abstract}


\section{Introduction}
This document is designed to accompany the software we have released
to compute the \WNF\
field\footnote{http://icc.dur.ac.uk/Panphasia.php}.  We assume that
the reader broadly understands what this software is for and has read
the introduction of \cite{Jenkins_13} and at least perused the rest of
that paper.  Our goal for this document is to provide sufficient
technical information and guidance to the reader to make using
the code relatively straight forward.  We are placing this document on
the arXiv as we think it will likely need to be updated from time to
time -- as will the software. We would welcome collaborators interested
in modifying or rewriting the code to improve the performance and to
provide support for languages other than fortran. We welcome comments
on how to improve this document.  We would be happy to add additional
authors to this document and to expand the acknowledgements section in
future versions.

 It is not necessary to follow the intricate internal workings of the
code to compute \WNF\ in order to use it.  We have provided the user
essentially with a function to compute the properties of
\WNF\ that can be treated as a black box. It is particularly simple to add
\WNF\ phases to an initial conditions code used to make ICs for
large-scale structure simulations - what we call `cosmological initial
conditions'.  For this application the user code only needs to call
two \WNF\ subroutines: one subroutine called once to initialise the
phases, and a second subroutine called repeatedly to evaluate the
field over a three dimensional cubic grid.

 We provide a set of subroutines with greater functionality for users
interested in making resimulation initial conditions. We describe a
method and provide reference calculations and examples in
\cite{Jenkins_13} to help with this.  ARJ would be happy to provide
advice on making resimulation initial conditions using \WNF.

As well as providing the code to compute \WNF, we have also included
several demonstration codes. These include a version of a public
serial code that has been modified so that it can make cosmological
initial conditions with \WNF\ phases. 

 In Section~2 we give an overview of \WNF\ and how it can be used
to publish phase information.  In Section~3 we give an overview
of the main subroutines we provide for the user for different
types of initial condition.  In Section~4 we describe the 
software package itself.   In Section~5 we present a few basic
tests of Gaussianity of the code.  In Section~6 we describe
the modifications we make to a public serial initial
conditions code so that it can make cosmological initial
conditions using \WNF\ phases.  We list and describe all
the subroutines in the appendices.

\section{Overview of PANPHASIA}
  \WNF\ is a very large predefined discrete realisation of a Gaussian
white noise field with a hierarchical structure based on an octree
geometry with 50 octree levels fully populated.  It is far too large
to be written down in entirety, but any part of \WNF\ can be evaluated
by calling a function we provide to compute the field.  This
function is serial and requires only a small amount of memory. 
A parallel code can simply hold as many private instances of the
function as needed.

 In \cite{Jenkins_13} we developed a text string, called a phase
descriptor that encapsulates the phase information for any given
simulation volume. Publishing the phase descriptor in a paper allows
others in principle to set up the same simulation volume and even
resimulate objects from that volume with no ambiguity as to the
phases\footnote{It is also necessary the paper gives the
cosmological parameters and linear power spectrum. This
is common practice already. For resimulations the
locations of regions of objects of interest must also be supplied.}.  
The phase descriptor specifies the phases for that volume on
all physical scales (down to the CDM free streaming scale if needs
be). 

To explain the format of the phase descriptor we need first to define
a set of coordinates to be used to refer to octree cells.  Any cell in
an octree at a level $l$ of the tree can be labelled using three
Cartesian integer coordinates each in the inclusive range of $0$ to 
$2^l-1$, where (0,0,0) corresponds to a corner cell situated at a
particular corner of the root cell of the octree.  We will call these
coordinates the absolute coordinates of the cell at level $l$.  We use
these absolute coordinates to define the phase descriptor.

\noindent A typical descriptor looks like this:

\vspace{5mm}
[Panph1,L11,(200,400,800),S3,CH439266778,MW7]
\vspace{5mm}

\noindent The descriptor uses five integers to define a cubic region
made of whole cells within the root cell of the octree.  The first of
these integers (11) specifies the shallowest level of the octree
needed to define the cube.  The next three integers (200,400,800) are
the absolute coordinates that label the octree cell at level 11 that
lies at the corner of the selected cube closest to the coordinate
origin at (0,0,0).  The next integer (3), gives the size of the cube
in units of octree cells at level 11.  Thus the cell furthest from the
coordinate origin in the cube has absolute coordinates (202,402,802).

The large integer (439266778) is a check digit that is computed from the
properties of \WNF\ at the location of the cube (full details are
given in \cite{Jenkins_13}, appendix~B).  The check digit provides a
way to detect for example a typo in a published phase descriptor. As
such a mistake will completely change the phase information for the
volume it is important to know that is there.  The last string, `MW7',
of the descriptor is a human readable name for the phases.  In this
case it names the phases used for Virgo Consortium's Millennium
simulation run with WMAP7 cosmological parameters. The name can be up
to 20 characters long with no spaces.

 Strictly speaking the phase descriptor defines a cubic 3-torus within
\WNF. This is because the phases are defined by the Gaussian white
noise field within this cube assuming periodic boundary conditions.
In general a 3-torus could be a cuboid, but the software we provide
assumes a cubic 3-torus. This is by far the most common boundary
condition used for cosmological simulations. If there is a demand for
support for a more general cuboid then let ARJ know.

 Once a 3-torus has been selected by using a phase descriptor there is
no need to use absolute coordinates to specify the octree cells within
it.  Instead our software uses relative Cartesian coordinates to label
cells in the 3-torus.  The user can choose the position of the origin
for these relative coordinates.  The relative coordinates are
non-negative integers, and obey the periodic boundary conditions of
the 3-torus.

 Finally, we provide a subroutine to generate new descriptors as it is
non-trivial to generate phase descriptors by hand because of the need
to compute the check digit.

\section{Overview of the software from the point of use}
 The software provides a set of subroutines that can be called
by the user's code.  There are two types of application that
we envisage and it is useful to consider these separately:
\begin{itemize}
\item Initial conditions for cosmological simulations.  These are for
uniform mass resolution simulations of periodic cubic volumes (or
3-tori) as typically used for studying large-scale structure.
\item resimulation initial conditions. These require multi-mass
particle distributions to focus the computational effort on a
sub-region of some larger simulation. This larger simulation
volume is a cubic 3-torus.  The phases used for any resimulation
are implicitly defined by the phase descriptor for the 3-torus
as a whole.
\end{itemize}
 The \WNF\ and pseudorandom number subroutines and functions are
described in greater detail in the appendices.

\subsection{Cosmological simulations}
 There are two \WNF\ subroutines that need to be called by the
user's program.  There is an initialisation routine called, {\tt
start\_panphasia}, that takes as input a phase descriptor and a grid
size which is an integer.  There is an evaluation subroutine called
{\tt panphasia\_cell\_properties} that returns the values of the \WNF\
field for a given cell position.

 The value of the grid size implicitly decides which level in the
octree is used by {\tt panphasia\_cell\_properties}.  Taking the phase
descriptor given previous section as an example, a value of 3,
corresponding to the side-length of the cube at level 11, would select
level 11. In general a value of $3\times2^n$ selects level $11+n$.
For any other value of the grid size an error message would be
produced, as these are the only values that allow a one-to-one
correspondence between the octree cells and the grid cells of a three
dimensional cubic mesh spanning the periodic simulation
volume. Without a one-to-one correspondence it is not possible to
accurately reconstruct the phase information.  

Because the grid size is quantised in this way it may be wise when
defining a new descriptor to choose the number of octree cells on a
side with care. Another factor to consider when choosing this value
are possible restrictions due to the parallelisation of the initial
conditions code.  For example a code may distribute a cubic mesh
across processors so that each processor has a whole number of planes
(e.g. because it uses {\sc FFTW}\ 2.1.5). In such a case it may be
desirable when running the code for the number of Fourier planes to be
exactly divisible by the number of mpi processes running.  The number
of mpi processes in turn may be determined by the number of cores per
node, at least when the code is making optimal use of the hardware.
The numbers of cores per node varies between machines and over time,
but common values are of the form either $2^n$ or $3\times2^n$, where
$n$ is a non-negative integer.  Choosing a side-length of 3 rather
than 1 for the phase descriptor provides more flexibility and 
we recommend this.

 The subroutine {\tt panphasia\_cell\_properties} returns the
properties of a single octree cell.  The routine uses a set of
relative coordinates to locate a cell within the 3-torus defined by
the phase descriptor and only returns the properties of the cell in
this prescribed domain. When the {\tt start\_panphasia} initialisation
routine is called the relative coordinate origin (0,0,0) corresponds
to the corner cell of the 3-torus that is closest to the absolute
origin of \WNF.

The {\tt panphasia\_cell\_properties} subroutine returns nine values
for each cell.  The first eight are the coefficients of the Legendre
block functions for that cell.  The Legendre block functions are
defined in Section~3.1 of \cite{Jenkins_13}.  The eight values are
ordered as: $p_{000}$, $p_{001}$, $p_{010}$, $p_{011}$, $p_{100}$,
$p_{101}$, $p_{110}$, $p_{111}$. The ninth value is independent and
not part of the \WNF\ field, although as described in Section~5.2 of
\cite{Jenkins_13} it may be used in making cosmological initial
conditions.  These nine values are independent Gaussian pseudorandom
numbers drawn from a distribution with zero mean and unit variance.

The {\sc cps}\ code, described in Section~\ref{CPS}, provides some
example code that shows how to combine the nine values returned by
{\tt panphasia\_cell\_properties} into a single field in $k$-space
that can then be used to make cosmological initial conditions in the
standard way using Fourier methods.

\subsection{Resimulation simulations}
We provide a separate set of subroutines with greater functionality
for making resimulation initial conditions.  Making such ICs typically
requires computing the phase information over a series of nested
grids or refinements about the region of interest.  The extra
capabilities include the ability to place the origin of the relative
coordinate system at any place within the 3-torus.  These relative
coordinates obey periodic boundary conditions with respect to the
3-torus and this feature saves the user some bookkeeping. The user can
also choose which levels of the octree contribute octree functions to
the returned Legendre coefficients of each cell. This gives the
user control on how to partition the octree basis functions between
the nested grids where they overlap.

The initialisation subroutine in this case is {\tt
set\_phases\_and\_ref\_orign}.  The first argument is again the
phase descriptor.  The second is the level of the octree the user
requires the \WNF\ field to be evaluated. In addition the user can 
choose the position of the origin used for the relative coordinates within the
3-torus.

For resimulation initial conditions the field is evaluated using a
function {\tt adv\_panphasia\_cell\_properties} which is a generalised
version of the subroutine {\tt panphasia\_cell\_properties}.  The
value of the field returned is relative to the origin of the refinement
specified by the {\tt set\_phases\_and\_refinement} subroutine. 

 The {\tt adv\_panphasia\_cell\_properties} subroutine allows the user to
choose the range of levels of the octree where the octree basis
functions are included in the calculation of the returned octree
basis function coefficients.  The cell properties are returned
in an identical structure to that used by {\tt panphasia\_cell\_properties}.
 Because range of octree basis functions can be restricted, the the values
returned by this function are no longer necessarily drawn from
a Gaussian distribution of zero mean and unit variance.

 We also provide a subroutine called {\tt layer\_choice}.
This subroutine is used by the \ICgen\ code when making resimulation 
initial conditions to determine which octree cells are placed
on which refinement. See Section~5.3 of \cite{Jenkins_13} and the appendix of
this paper for more details.

\section{Contents of the software package}
 The software is packaged as a gzipped tar file called {\tt PANPHASIA\_code\_V1.000.tar.gz}.
It unpacks to give a top level directory {\tt PANPHASIA\_public\_code} which contains
a short {\tt README} file and three subdirectories: {\tt src, demo, cps}. This
{\tt README} file will be expanded in future versions to include details of
revisions to the software.

The directory {\tt src} contains the source code for computing
\WNF. There are two source files: {\tt panphasia\_routines.f} written by ARJ;
and {\tt generic\_lecuyer.f90} the implementation of \MRG\
pseudorandom number \citep{Lecuyer93} written by SPB.  The subroutines
in {\tt panphasia\_routines.f} and {\tt generic\_lecuyer.f90}  are listed in Tables~\ref{panphasia_subs}
and \ref{mrg_routines}. Detailed descriptions of these subroutines are given in the
appendices. 

The directory {\tt demo} contains a program file ({\tt main.f}) and an example
makefile. The demonstration code has five different modes. These are
described in detail, with examples of the code output provided in the ascii file
{\tt README\_demo}. We recommend the user runs these demonstrations
and checks the outputs against those illustrated in this {\tt README}\ file.
\begin{itemize}
\item  a subroutine to compute a random descriptor. This provides a convenient way
to create descriptors for new simulations.
\item a subroutine to calculate the zeroth and first moments of the overdensity
   field defined by \WNF\ over a cuboidal region specified by the user. 
   This provides a test of the code to make sure it is
    correctly computing the field by comparing the output to that given in the
   {\tt README}\ file.
\item  a subroutine ({\tt random\_summation\_demo}) intended to act as a warning about the
code performance when using a close to random access pattern to evaluate \WNF.
\item a subroutine ({\tt raster\_summation\_demo} which shows the best and worst
     raster scan patterns to use for evaluating \WNF.
\item a subroutine ({\tt fast\_summation\_demo}) that shows a particularly efficient access
  pattern suitable suitable for adding to an initial conditions code. We recommend
  this access pattern is used in parallel applications. 
\end{itemize} 

 The directory {\tt cps} contains a version of the serial cosmological
initial conditions code \cite{CPS_06}. This code has been modified so
that it can read in a \WNF\ descriptor. Two example run scripts are
provided. See Section~\ref{CPS} for more details.

 The underlying structure of \WNF\ is completely specified by a large
number of predetermined integer values (in fact $>10^{46}$). The
values of the field however are floating point numbers that are
derived from these integers.  The process of generating the floating
point values is affected by double precision rounding errors.  The
exact values of the output floating point numbers will depend on
details that effect the order of operations such the particular
compiler and the compiler flags used.  The differences arising from
alternate choices of compilers and optimisation flags are very small
in our experience and not a significant concern for the purposes of
making cosmological or resimulation initial conditions. We find the
speed of the code is sensitive to the compiler flags, and is
significantly speeded up with the most aggresive
optimisation. However, we advise the user check the effects of
different compiler flags themselves. Running the demonstration codes
and comparing the output produced with the output shown in {\tt
README}\ file provides a way to do this. 

\begin{figure}
\resizebox{\hsize}{!}{
\includegraphics{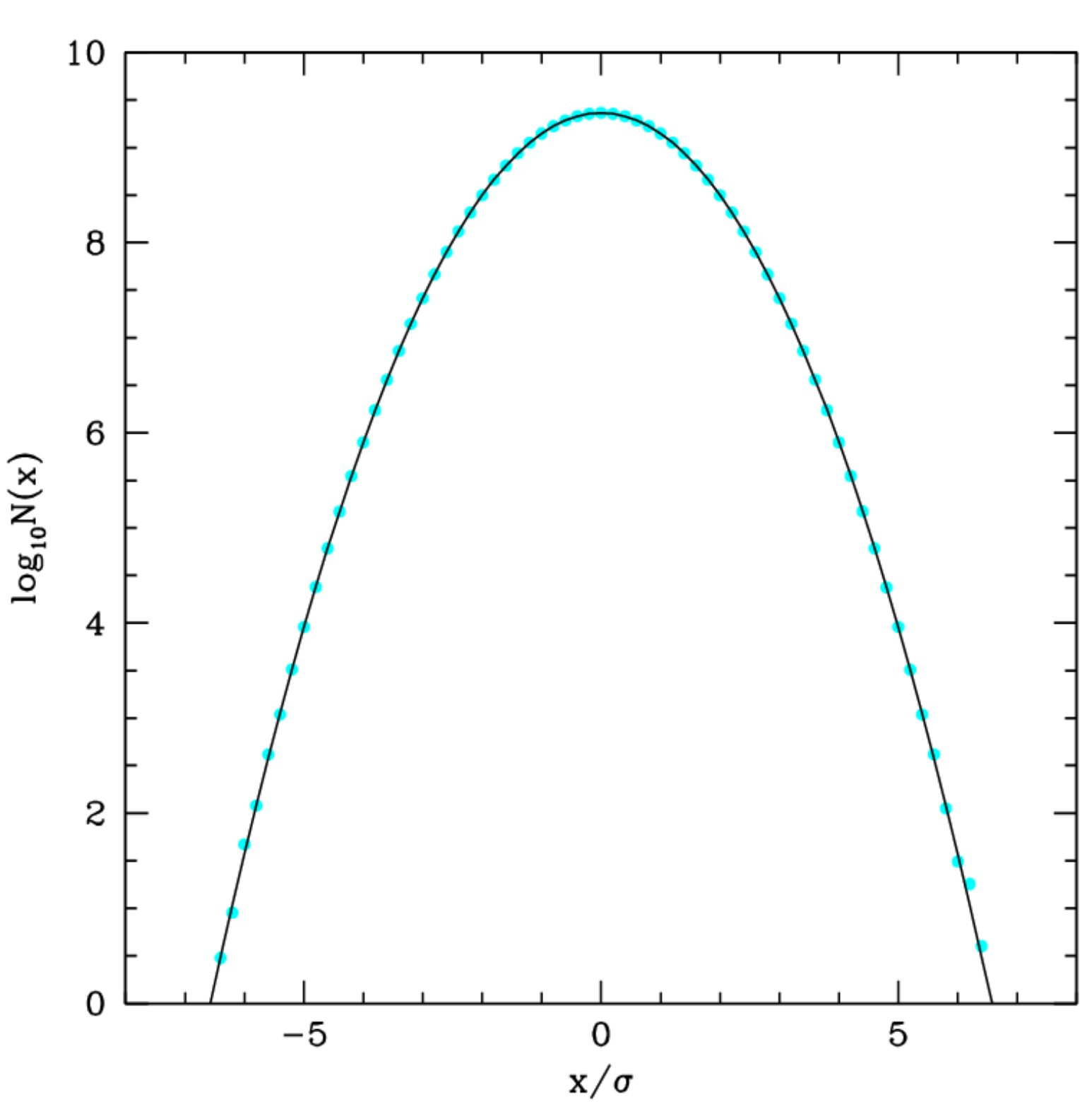}}
\caption{The one-point distribution of the $p_{000}$ Legendre
coefficient evaluated over a $3072^3$ grid cells for a
randomly chosen region in \WNF. The points show
the number of grid cells with values falling within bins of 
width $0.2\sigma$ centred on zero. The curve shows the
expected Gaussian distribution.  The agreement is very good.
The value of $\sigma$ for the one-point function is unity.}
\label{one_point}
\end{figure}

\begin{figure}
\resizebox{\hsize}{!}{
\includegraphics{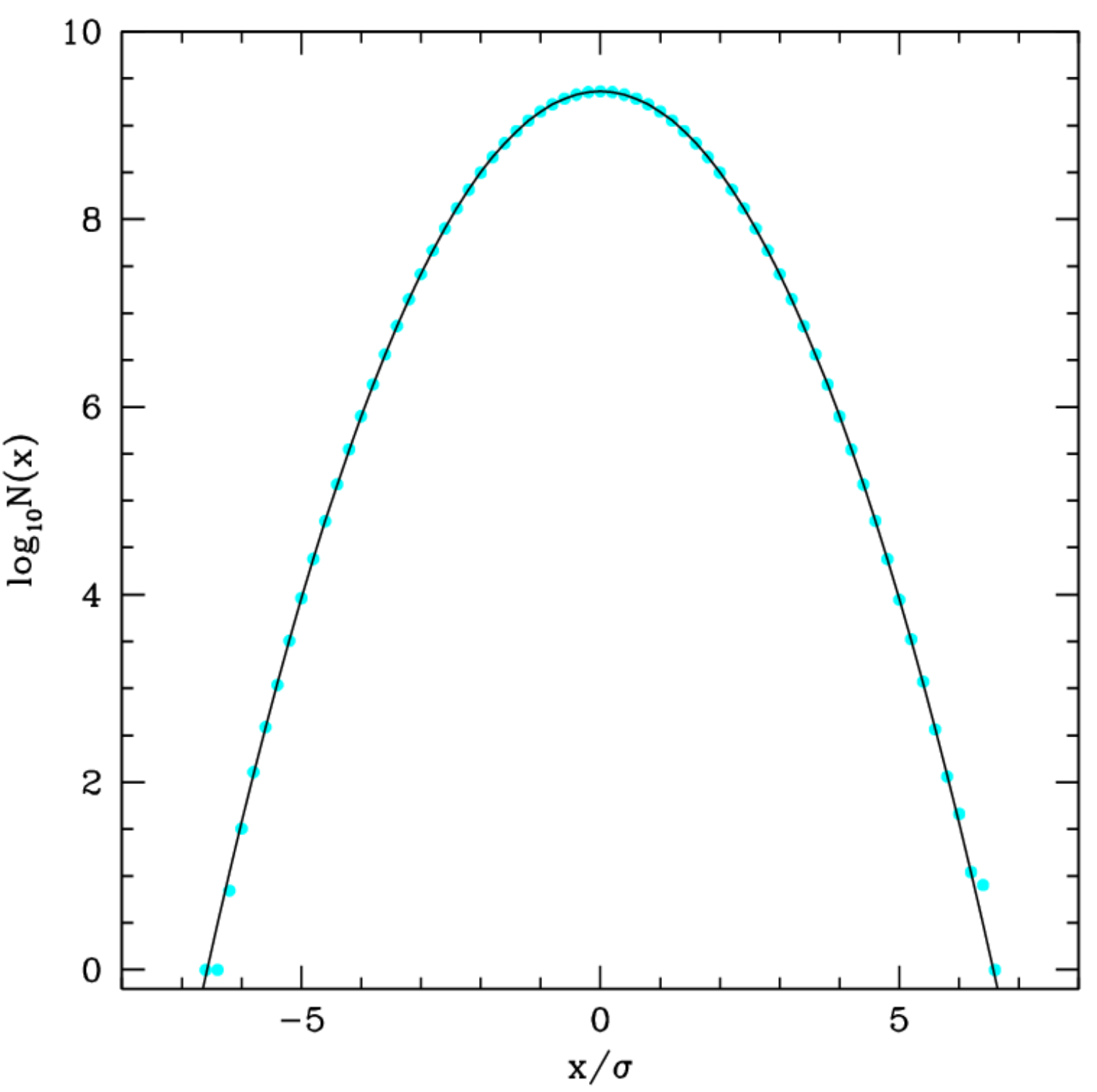}}
\caption{The distribution of values of the two-point cross-correlation function
evaluated between the $p_{000}$ and $p_{001}$ Legendre coefficients 
over a $3072^3$ grid cells for the same region as
Figure~\ref{one_point}. The points show the number of grid cells at
non-zero lag with values falling within bins of width $0.2\sigma$
centred on zero. While the plot looks very like Figure~\ref{one_point}
the value of $\sigma = 1/3072^{3/2} = 5.88\times10^{-6}$ so the fields
defined by the $p_{000}$ and $p_{001}$ Legendre coefficients are
essentially uncorrelated.  The same is true of the other 35 possible
cross-correlations and the 9 autocorrelation functions at non-zero
lag.}
\label{two_point}
\end{figure}

\section{Test of the PANPHASIA code}
  We present some basic tests of the \WNF\ code for a region specifed
by a phase descriptor. We evaluate the field over a large three
dimensional grid. For each grid point the {\tt
panphasia\_cell\_properties} routine returns nine values.

 We expect that the one-point distribution of each of these nine
values should accurately follow a Gaussian distribution with zero mean
and unit variance.  The two-point correlation function evaluated over
the grid for each set of values should be very close to zero except at
zero lag.  The cross-correlation function between any two sets of
values should be close to zero at all lags.

 While the ensemble averaged auto and cross correlation functions
should in general be zero,  for a finite sample the most we can
expect is that they are very small.  For a sample of $N$ grid points
we expect the one-point distribution of the cross-correlation functions,
and the autocorrelation functions at non-zero lag,  to be
accurately Gaussian with a mean of zero and a small variance of $1/N$.

 For our tests we use a $3072^3$ grid, which enables the distribution
to be well measured within the range of about $\pm6$ standard deviations of
the expected Gaussian distributions. For the purposes of the
test we generated a random phase
descriptor:

\vspace{5mm}
[Panph1,L40,(592773559564,63351353918, \\
901943199905),S3,CH1532937555,Test\_example]
\vspace{5mm}

We have tested other descriptors and get similar results. It is not
however practicable to test a significant fraction of the \WNF\ field
except at low levels of the octree which we do not expect to be
used in entirety to make initial conditions anyway.

 Figure~\ref{one_point} shows the one-point distribution function for the $p_{000}$
Legendre basis function. Plots for the other eight values look very similar.
The level of agreement is excellent even in the tail of the
distribution.  Taking the 61 bins encompassing the range -6.1 to 6.1
$\sigma$ of the distribution about zero we calculated a $\chi^2$ value
from the difference of the measured and expected numbers in each bin
and obtained a value of 68.2. This is close to the expected value of
61. We conclude the one-point distribution is as expected very close
to Gaussian.

We computed the cross and autocorrelation functions of the nine fields
using Fourier methods.  For all 36 cross-correlation functions,
between any two pairs of the nine values returned by {\tt
panphasia\_cell\_properties} we find the cross-correlation be very
small at all lags as expected. The values follow a Gaussian
distribution with a variance of $1/3072^3$.  From the absence of any
outliers beyond $7\sigma$ we can set an firm upper limit to the maximum
correlation at any lag to be $7/3072^{3/2} =
4.1\times10^{-5}$.  For the 9 autocorrelation functions we obtain
similar results, after exempting the very strong correlation
expected at zero lag. Figure~\ref{two_point} shows the distribution of
values for the cross-correlation function between the $p_{000}$
and $p_{001}$ Legendre blocks.

 In conclusion we find no evidence for unexpected correlations in
these limited test. Passing these tests is necessary, but not
sufficient to demonstrate that there are no unwanted correlations in
our Gaussian white noise field.  We strongly encourage others to test the
randomness of \WNF\ and report the results to us.  We hope to include
the results of more tests in future versions of this document.

 As we discussed in \cite{Jenkins_13} it is certain that the
pseudorandom number generator that we use will fail randomness tests
for a sufficiently large sample.  Very small deviations from
randomness are not obviously a concern for making cosmological initial
conditions.  If larger deviations are found in other test then these
need to be quantified to decide their significance for structure
formation simulations.  In the worst case it may be necessary to fix
the problem - which is a bad from the point of view of publishing
phase information as a new phase field would result. The phase
descriptor format can accomodate this - although we hope this will not
prove necessary.

 Deviations from randomness can arise not just from the pseudorandom
generator but from mistakes in the software. Indeed the tests in this
section are aimed at primarily identifying code errors.  We find by
deliberately breaking the code that small errors for example in the
coefficients defining the octree basis functions are readily detected.
The accidental use of the pseudorandom numbers more than once also can
be detected.  ARJ tested the pseudorandom number routines written by
SPB by writing his own generator. This allowed testing that the
pseudorandom sequence was correctly being mapped to the octree.

\section{An example of adding PANPHASIA to code\label{CPS}}
 As part of the code distribution we provide a modified version of
a serial initial conditions generator that accompanied the 
paper \cite{CPS_06}.  The code for this is in the {\tt cps}
directory. We will refer to the code as the {\sc cps} code.

 We have made fairly minimal alterations to the original code. The
main change we have made is to separate the part where the phases are
set, from the part where the Zeldovich displacements are computed in
$k$-space.  This makes it easy to put a conditional statement in
the code the provides the user with a choice to use original 
method of setting up the phases,  or phases derived from an input \WNF\
phase descriptor.

 The main addition to the code is a new subroutine called {\tt initialise\_wnf} 
that calculates the \WNF\ phases.  This subroutine is placed in a separate file called
{\tt set\_up\_wn\_field.f} and has extensive comments.  The {\tt initialise\_wnf}
subroutine is called once from the main program. 

The {\sc cps}\ code makes {\sc 2lpt} initial conditions. To do this it
needs to store six values per grid cell. The {\tt initialise\_wnf}
subroutine is able to use these grids as temporary storage. As a
result the addition of \WNF\ phases to this code does not increase the
memory requirements.  It is however necessary to call the subroutine
that returns the properties of \WNF\ twice per grid cell.  This is
because the routine returns nine numbers per grid cell - only six of
which can be stored at one time.  The {\tt initialise\_wnf} subroutine
combines these nine numbers into a single $k$-space field that
reproduces the \WNF\ phases in a two stage process: Stage 1 -- taking
first five values on five separate grids and combining them into a
single grid in $k$-space; Stage 2 -- assigning the remaining four to
four new grids and combining these again in $k$-space together with
the single grid from stage 1 into a final single field.  This field is
then passed back to the main program and used to generate the initial
conditions in the same way as was done in the original code.

 Based on limited experience we have found that the process of adding
a new method for setting phases to an initial conditions code can
prove tricky in practice. Even when the new method is almost working
the output initial conditions may still bare no obvious obvious
resemblance to what is expected!  It is advisable to make initial
conditions at high redshift so that the output is a close to a linear
function of the input phases.  Because the process of making the
initial conditions is essentially linear it is a good idea to focus on
getting the phase and amplitude of single Fourier mode right before
attempting to get multiple modes to work. The step from getting a
single mode to multiple modes working is not always completely
simple.

Our first attempt to add \WNF\ phases to the {\sc cps} code was at
first only partially successful.  After first focusing on getting the
normalisation of the waves correct, we observed that individual modes
in the output initial conditions had phases that were either plus or
minus $\pi/2$ in phase different from that given by \WNF.  The reason
for these shifts is due to the absence of factors of $i$ in the
computation of the Zeldovich displacements in the {\sc cps}\ main
program.  As our aim was to change the original {\sc cps}\ code as
little as possible, we corrected these unwanted phase shifts in the
output initial conditions by adding a series of $\pi/2$ phases changes
to each mode at the end of the {\tt initialise\_wnf} subroutine. These
shifts simply compensate for what is essentially a `feature' of this
particular code.

 Once this was done then we then found good agreement, at close to
single precision round-off, between the {\sc cps} code Zeldovich
initial conditions and Zeldovich initial conditions produced by the
\ICgen\ code. To get this level of agreement it was necessary to use
slightly different values of the $\sigma_8$ parameter in the two codes
because the power spectrum normalisation in the {\sc cps} code is not
as accurate as in \ICgen. It is also crucial to be sure that exactly
the same modes are being set by both codes.  Differences in the modes
used between initial conditions code usually arise because of
different ways of imposing a high $k$ cut-off are used. The modified
{\sc cps} code uses the same sharp spherical cut-off as \ICgen\ for
\WNF\ phases.

 We have included two run scripts for the {\sc cps} code. These are:
{\tt run\_zeld.sh} which produces the initial conditions that we used
in our testing against the \ICgen\ code, and {\tt dove\_example384.sh}
which generates cosmological initial conditions for the {\sc dove}
volume used in the test simulations in Section~6 of \cite{Jenkins_13}.  Running
these {\sc dove} initial conditions with the Gadget-2 code \citep{Springel_05} to
redshift zero we are able to locate the Milky-way mass halo studied in
\cite{Jenkins_13} at the expected location.  The halo is reproduced
with only about 2000 particles at this resolution. We do find a halo
close to the correct position and mass.  This is very unlikely to be a
coincidence. The position of the halo is reproduced to a few tens of
kiloparsecs and mass to a few percent of the values given in
\cite{Jenkins_13}. We provide the Gadget-2 parameter file {\tt
Gadget2/dove\_cps\_384.param} that we used to run this simulation in our
software package.

\section*{ACKNOWLEDGEMENTS}
I would like to thank Martin Crocce for permission to include the
modified 2lpt serial code with our software. Anyone making use of this
code should reference \cite{CPS_06} and \cite{Jenkins_13} in any
published papers that result. Thanks also to Nigel Metcalfe to setting
up the \WNF\ download page.

\bibliography{wn_doc}

\begin{thebibliography}{4}
\expandafter\ifx\csname natexlab\endcsname\relax\def\natexlab#1{#1}\fi

\bibitem[{{Crocce} {et~al.}(2006){Crocce}, {Pueblas}, \&
  {Scoccimarro}}]{CPS_06}
{Crocce} M., {Pueblas} S., {Scoccimarro} R., 2006, MNRAS, 373, 369

\bibitem[{{Jenkins}(2013)}]{Jenkins_13}
{Jenkins} A., 2013, ArXiv e-prints

\bibitem[{L'Ecuyer {et~al.}(1993)L'Ecuyer, Blouin, \& Couture}]{Lecuyer93}
L'Ecuyer P., Blouin F., Couture R., 1993, ACM Transactions on Modeling and
  Computer Simulation, 3, 87

\bibitem[{{Springel}(2005)}]{Springel_05}
{Springel} V., 2005, MNRAS, 364, 1105

\end{thebibliography}

\appendix

\section{Description of the panphasia\_routines code}\label{pan_code}
In this appendix we list all of the subroutines in the file
{\tt panphasia\_routines.f} and describe their inputs and outputs
and what they do. We list them in the order they occur in the
file. 
\subsection{}
\begin{proto}
SUBROUTINE start\_panphasia(descriptor,Ngrid) \\
CHARACTER*100, INTENT(IN)::string \\
INTEGER, INTENT(IN)::Ngrid 
\end{proto}

This subroutine needs to be called be a user code making cosmological
initial conditions.
The {\tt descriptor} is a character string holding a \WNF\ phase
descriptor. {\tt Ngrid} is the size of a cubic grid that spans the
3-torus defined by the descriptor. The value of {\tt Ngrid} must be a
power of two times the integer value prefixed with an `S' in a given
descriptor.  This restriction is necessary to ensure a one-to-one
correspondence between octree cells and the grid that it is being
assigned.
\subsection{}
\begin{proto}
SUBROUNTINE set\_phases\_and\_rel\_origin(descriptor, \\
lev, ix\_rel,iy\_rel,iz\_rel)\\
CHARACTER*100, INTENT(IN)::descriptor \\
INTEGER*8, INTENT(IN)::ix\_rel,iy\_rel,iz\_rel \\
CHARACTER*20, INTENT(IN)::phase\_name
\end{proto}
A more general initialisation routine. The routines takes a \WNF\
phase descriptor as a first argument, but requires the user to
explicitly set the level of the octree that will be used to return the
properties of the field.  The parameters {\tt ix\_rel,iy\_rel,iz\_rel}
define the origin of the relative coordinate system, at level {\tt lev} 
of the octree, used to evaluate
the field. For resimulations this origin is typically placed at a
corner cell of a cubic refinement.  The values of {\tt
ix\_rel,iy\_rel,iz\_rel} are constrained to lie within the 3-torus.
The relative coordinates obey the periodic boundary conditions of the
3-torus and each component must be non-negative.
\begin{table}
{
\begin{tabular}{|l|l|l|l|}
\hline
Label   &\WNF\ subroutines               &Called by  &Calls             \\
\hline
A1   &{\tt start\_panphasia}                &C            & A2,11      \\
A2   &{\tt set\_phases\_and\_rel\_origin}    &R            & A3,11,13  \\
A3   &{\tt initialise\_panphasia}           &A1,2        & --          \\
A4   &{\tt panphasia\_cell\_properties}     &C,D          &A5           \\
A5   &{\tt adv\_panphasia\_cell\_properties}&R,A4         &A6           \\
A6   &{\tt return\_cell\_props }            &A5           &A7,8,10    \\
A7   &{\tt evaluate\_panphasia }            &A6           &--           \\
A8   &{\tt reset\_lecuyer\_state}           &A6           &A9           \\
A9   &{\tt advance\_current\_state}         &A8           &--           \\
A10   &{\tt return\_gaussian\_array}        &A6           &--           \\
A11  &{\tt parse\_descriptor}               &R,A1,2,13  &--           \\
A12& {\tt compose\_descriptor}              &A14,18      &--           \\
A13& {\tt validate\_descriptor}             &A2,14,18 &A11          \\
A14& {\tt generate\_random\_descriptor}     &D            &--           \\
A15& {\tt demo\_basis\_function\_allocator } &--           &A16,17      \\
A16& {\tt layer\_choice}                    &R,A15        &A17          \\
A17& {\tt inref }                           & A16        &--            \\
A18& {\tt set\_local\_box}                   & D           &A3,12,13   \\
\hline
\end{tabular}
}
\caption{\WNF\ subroutines and their dependencies. The label `C'
represents an intial conditions code for making cosmological initial
conditions. Such a code would only need to call subroutines A1 and A4
directly.  The label `R' represents an initial conditions code for
making resimulation initial conditions (e.g. \ICgen).  The label `D'
represents one or other of the demonstration subroutines that are
packaged with the software.  The labels refer to the appropriate
subsections in appendix~\ref{pan_code} where the routines are
described in detail.}
\label{panphasia_subs}
\end{table}
\subsection{}
\begin{proto}
SUBROUTINE initialise\_panphasia \\
\end{proto}
Not called directly be the user. This subroutine initialises both \MRG\ generator and \WNF. 
\subsection{}
\begin{proto}
SUBROUTINE panphasia\_cell\_properties(ix,iy,iz, \\
cell\_prop) \\
INTEGER, INTENT(IN)::ix,iy,iz \\
REAL*8, INTENT(OUT)::cell\_prop(9)
\end{proto}
A basic function to evaluate the Legendre basis function coefficients
suitable for making cosmological initial conditions. The cell location
is given by relative integer coordinates {\tt (ix,iy,iz)}. These
relative coordinates are implicitly determined by an earlier call to
{\tt start\_panphasia}. The values of the expansion coefficients are
returned in the structure, {\tt cell\_prop(9)}.  The first 8 values of
cell\_prop are the Legendre block expansion coefficients of \WNF\ for
that particular cell.  The ninth entry is an independent pseudorandom
Gaussian number that is not part of \WNF. In \cite{Jenkins_13} this
ninth value is used to construct an independent Gaussian field to
ensure that cosmological initial conditions are isotropic at small
scales.

The first eight entries for the {\tt cell\_data} array are the
Legendre block coefficients ordered as: $p_{000}$, $p_{001}$,
$p_{010}$, $p_{011}$, $p_{100}$, $p_{101}$, $p_{110}$, $p_{111}$. 
All nine numbers are drawn from a Gaussian distribution with zero
mean and unit variance.
\subsection{}
\begin{proto}
SUBROUTINE adv\_panphasia\_cell\_properties(ix,\\
iy,iz,layer\_min,layer\_max,indep\_field,\\
cell\_prop) \\
INTEGER, INTENT(IN)::ix,iy,iz \\
INTEGER, INTENT(IN)::layer\_min,layer\_max,\\
INTEGER, INTENT(IN)::indep\_field \\
REAL*8, INTENT(OUT)::cell\_prop(9)
\end{proto}
This is a more sophisticated version of the {\tt
panphasia\_cell\_properties} subroutine described above.  It has three
extra arguments that can be used to control the range of levels of the
octree basis functions used to compute the Legendre block
coefficients. This extra functionality is needed for making
resimulation initial conditions.  As above the cell location is given
by relative integer coordinates {\tt (ix,iy,iz)}. These relative
coordinates are determined by an earlier call to either {\tt
start\_panphasia} or more usually {\tt set\_phases\_and\_rel\_origin}. The values of
the expansion coefficients are returned in the structure, {\tt
cell\_prop(9)}, and depend on exactly which octree basis functions are
included.  The octree functions selected is controlled by the
parameters {\tt layer\_min} and {\tt layer\_max} which determine which
levels of the octree are populated with the octree basis functions.
For cosmological initial conditions the natural choice for these are 0
and {\tt lev} so that all the possible octree functions are used.

 The subroutine returns the Legendre block expansion coefficients as
an array of dimension 9 in the same order as {\tt
panphasia\_cell\_properties}.  The ninth entry is not properly part of
\WNF\ and the output value also depends on the value of the input
integer {\tt indep\_field}. If this integer is 0 then the value
returned is also zero. If it 1 then the returned value is an
independent Gaussian pseudorandom number drawn from a distribution of
mean zero and unit variance.
 A value of -1 for {\tt indep\_field} results in the independent field
being returned and the Legendre basis coefficients all set to
zero. This is just testing purposes.
\subsection{}
\begin{proto}
SUBROUTINE return\_cell\_props(lev\_input,ix\_half,\\
iy\_half,iz\_half,px,py,pz,layer\_min  \\
layer\_max,indep\_field,cell\_data)\\
INTEGER, INTENT(IN)::lev\_input, ix\_half, \\
INTEGER, INTENT(IN)::iy\_half, iz\_half\\
INTEGER, INTENT(IN)::px,py,pz, layer\_min\\
INTEGER, INTENT(IN)::layer\_max\\
REAL*8,  INTENT(OUT)::cell\_data(9,0:7) \\
\end{proto}

Not called by the user.
 This routine manages the stored octree. The code holds at most eight
octree cells at each level of the tree all focused about a particular
spatial location within the root cell. If the user requests a value of
the field which lies outside of any of the known cells, the relevant
parts of the octree are rebuilt to include the new location. Some of
the information for the previous location is then lost.  A random
access pattern of the field therefore requires a lot of computation
and is best avoided.  If the value of the field has already been
computed and is currently being stored then the routine simply returns
the values and avoids doing a new computation.
\subsection{}
\begin{proto}
SUBROUTINE evaluate\_panphasia(nlev,leg\_coeff,\\
maxdim,gauss\_list,layer\_min,layer\_max,indep\_field,\\
icell\_name, cell\_data) \\
INTEGER, INTENT(IN)::nlev ,maxdim\\
REAL*8, INTENT(IN)::gauss\_list(*) \\
INTEGER, INTENT(IN)::layer\_min,layer\_max \\
INTEGER, INTENT(IN)::indep\_field \\
INTEGER, INTENT(IN)::icell\_name \\
REAL*8,  INTENT(INOUT)::leg\_coeff(0:7,0:7, \\
$\phantom{XXXXXXXXXXXXXXXXXXXX}$-1:maxdim) \\
REAL*8,  INTENT(OUT)::cell\_data \\
\end{proto}
Not called by the user.
 This routine calculates the Legendre basis function coefficients
for the eight child cells of a given parent octree cell. These
coefficients depend on both the Legendre basis function
coefficients of the parent cell, and the octree basis functions
belonging to the parent cell.  The routine combines the
information from these two sources of information to calculate the 
Legendre basis function coefficients of all eight child
cells.  

 This calculation is done following the definitions of the
octree basis functions given at the start of appendix~A
of \cite{Jenkins_13}. The subroutine is well commented.
\subsection{}
\begin{proto}
SUBROUTINE reset\_lecuyer\_state(lev,xcursor,\\
ycursor,zcursor)\\
INTEGER,   INTENT(IN)::lev \\
INTEGER*8, INTENT(IN)::xcursor,ycursor,zcursor
\end{proto}
Not called by the user.
 The code keeps a three dimensional `cursor' at each level of the octree
that holds the cells of interest.  This cursor is positioned to ensure
that the correct pseudorandom sequence is chosen for each octree basis
function.  
\subsection{}
\begin{proto}
SUBROUTINE advance\_current\_state(lev, \\
xcursor,ycursor,zcursor) \\
INTEGER, INTENT(IN)::lev \\
INTEGER*8, INTENT(IN)::xcursor,ycursor,zcursor 
\end{proto}
Not called by the user.
A low level routine that ensures that the relevant part of the
pseudorandom number sequence is used in each cell of the
subroutine {\tt return\_gaussian\_array} below.
\subsection{}
\begin{proto}
SUBROUTINE return\_gaussian\_array(lev,\\
ngauss,garray) \\
INTEGER, INTENT(IN)::lev, ngauss \\
REAL*8,  INTENT(OUT)::garray(0:ngauss-1) 
\end{proto}
Not called by the user.
This routine makes an even number, in practice in the code either
eight or sixty-four consecutive calls to the \MRG\ routine and uses a
Box-Muller transformation to produce eight or sixty-four pseudorandom
Gaussian variables with zero mean and unit variance.  The precise
method used is described in appendix~B of \cite{Jenkins_13}.
\subsection{}
\begin{proto}
SUBROUTINE parse\_descriptor(string,l,ix,iy,iz, \\
side1,side2,side3,check\_int,name) \\
CHARACTER*100, INTENT(IN)::string \\
INTEGER, INTENT(OUT)::l \\
INTEGER*8, INTENT(OUT)::ix,iy,iz \\
INTEGER, INTENT(OUT)::side1,side2,side3 \\
INTEGER*8, INTENT(OUT)::check\_int \\
CHARACTER*20, INTENT(OUT)::name 
\end{proto}
 This routine takes a phase descriptor as input and parses it into its
constituent parts and returns them.  If the descriptor is malformed it
will complain.  This routine only works for descriptor that defines a
cubic volume. Therefore the output values of the sides obey: {\tt
side1 = side2 = side3}.
\subsection{}
\begin{proto}
SUBROUTINE compose\_descriptor(l,ix,iy,iz, \\
side,check\_int,name,string) \\
INTEGER, INTENT(IN)::l \\
INTEGER*8, INTENT(IN)::ix,iy,iz \\
INTEGER, INTENT(IN)::side \\
INTEGER*8, INTENT(IN)::check\_int \\
CHARACTER*20, INTENT(IN)::name \\
CHARACTER*100, INTENT(OUT)::string
\end{proto}
This routines creates a phase descriptor from the inputs. It does not
calculate the check digit. To do this use {\tt validate\_descriptor}
described below to calculate the check digit, and then call {\tt
compose\_descriptor} again.
\subsection{}
\begin{proto}
SUBROUTINE validate\_descriptor(string, \\
MYID, check\_int) \\
CHARACTER*100, INTENT(IN)::string \\
INTEGER, INTENT(IN)::MYID \\
INTEGER*8, INTENT(OUT)::check\_int 
\end{proto}
This routine performs basic checks on a phase descriptor to see if it makes
sense. This includes making sure that the check digit is consistent
with the other information in the descriptor. If it finds an error the
subroutine it just stops. For the special case where the check digit
is -999, and the descriptor is otherwise correct the routine returns
the value of the check digit.
\subsection{}
\begin{proto}
SUBROUTINE generate\_random\_descriptor(string) \\
CHARACTER*100, INTENT(OUT)::string
\end{proto}
 This utility can create new descriptors for the user.  It
tries to choose them randomly from the \WNF\ volume. It
uses the unix\_timestamp, and some input user specific
information to help select the region.  

 The user is asked the side of the simulation in Mpc/h,
so that it can choose which level of the octree to select
as most appropriate.  This feature is intended to ration
the available \WNF\ volume so that the user is unlikely
to choose phases that have been used before.

 We would ask users to respect this scaling when generating
descriptors, unless there is a very good reason for not
doing so. ARJ will also start using this convention 
for new simulation volumes.
\subsection{}
\begin{proto}
SUBROUTINE demo\_basis\_function\_allocator\\
\end{proto}
 This code was written to illustrate the function of
the next two routines which implement the heuristic
scheme discussed in \cite{Jenkins_13} for deciding
where to place particular octree basis functions 
when making resimulation initial conditions.
\subsection{}
\begin{proto}
SUBROUTINE layer\_choice(ix0,iy0,iz0,iref, \\
nref,ix\_abs,iy\_abs,iz\_abs,ix\_per,iy\_per, \\
iz\_per,ix\_rel,iy\_rel,iz\_rel,ix\_dim, \\
iy\_dim,iz\_dim,wn\_level,,x\_fact \\
layer\_min,layer\_max,indep\_field)\\
INTEGER, INTENT(IN):: ix0,iy0,iz0,iref,nref \\
INTEGER*8, INTENT(IN)::ix\_abs(nref),iy\_abs(nref) \\
INTEGER*8, INTENT(IN)::iz\_abs(nref),ix\_per(nref) \\
INTEGER*8, INTENT(IN)::iy\_per(nref),iz\_per(nref) \\
INTEGER*8, INTENT(IN)::ix\_rel(nref),iy\_rel(nref) \\
INTEGER*8, INTENT(IN)::iz\_rel(nref)  \\
INTEGER, INTENT(IN)::wn\_level(nref),x\_fact \\
INTEGER, INTENT(OUT)::layer\_min,layer\_max \\
INTEGER, INTENT(OUT)::indep\_field
\end{proto}
This subroutine was used by the \ICgen\ code in \cite{Jenkins_13} to
implement the resimulation method.  Its function is to decide given a
set of nested refinements which octree basis functions should be
assigned to which refinement as a function of position on a given
refinement.  It returns the values of the {\tt layer\_min}, {\tt
layer\_max} and {\tt indep\_field} which can then be used as inputs to
the {\tt adv\_panphasia\_cell\_properties} subroutine. The values {\tt
ix0,iy0,iz0} are the relative coordinates of a cell in the {\tt iref}
refinement.  This refinement is one of {\tt nref}, all of which
have positions and sizes that are passed as the arguments {\tt ix\_abs(nref)} -
{\tt iz\_rel(nref)}.  The input parameter {\tt x\_fact} affects how
the octree cells are placed with respect to the refinement boundaries
and is introduced in \cite{Jenkins_13} in Section~5.3.
\subsection{}
\begin{proto}
SUBROUTINE inref(ixc,iyc,izc,isz,ir1,ir2, \\
nref,wn\_level, ix\_abs,iy\_abs,iz\_abs,\\
ix\_per,iy\_per,iz\_per,ix\_rel,iy\_rel,iz\_rel,\\
ix\_dim,iy\_dim,iz\_dim,x\_fact,interior,iboundary) \\
INTEGER, INTENT(IN) ::ixc,iyc,izc,isz,ir1,ir2,\\
INTEGER, INTENT(IN) ::nref,wn\_level\\
INTEGER*8, INTENT(IN)::ix\_abs(nref),iy\_abs(nref) \\
INTEGER*8, INTENT(IN)::iz\_abs(nref),ix\_per(nref) \\
INTEGER*8, INTENT(IN)::iy\_per(nref),iz\_per(nref) \\
INTEGER*8, INTENT(IN)::ix\_rel(nref),iy\_rel(nref) \\
INTEGER*8, INTENT(IN)::iz\_rel(nref),ix\_dim(nref)  \\
INTEGER*8, INTENT(IN)::iy\_dim(nref),iz\_dim(nref)  \\
INTEGER, INTENT(IN)::x\_fact \\
INTEGER, INTENT(OUT)::interior, iboundary 
\end{proto}
Not called by the user.
 This subroutine returns information on whether a cell
of interest is inside or outside of a refinement, and
if inside whether it is close to or far from the boundary.
\subsection{}
\begin{proto}
SUBROUNTINE set\_local\_box(lev,ix\_abs,iy\_abs,iz\_abs,\\
ix\_per,iy\_per,iz\_per,iz\_per,ix\_rel,iy\_rel,iz\_rel,\\
wn\_level\_base,check\_int,phase\_name) \\
INTEGER,   INTENT(IN)::lev,wn\_level\_base \\
INTEGER*8, INTENT(IN)::ix\_abs,iy\_abs,iz\_abs \\
INTEGER*8, INTENT(IN)::ix\_per,iy\_per,iz\_per \\
INTEGER*8, INTENT(IN)::ix\_rel,iy\_rel,iz\_rel \\
INTEGER*8, INTENT(IN)::check\_int \\
CHARACTER*20, INTENT(IN)::phase\_name
\end{proto}
This is a more general initialisation routine - it does not
take a phase descriptor as input, but some of the input parameters 
can be  dervied from a descriptor. The subroutine above
{\tt  parse\_descriptor} can be used to extract this
information from a descriptor. We use this subroutine in
one of the demonstration codes.

This subroutine allows the user to define both a 3-torus and a cubic
subregion within the 3-torus.  Calls to the {\tt
adv\_panphasia\_cell\_properties} subroutine use integer cooordinates which
are measured relative to this cubic subregion.

The variable {\tt lev} refers to the level of the octree required by
the user. The origin of the 3-torus is given in absolute coordinates
by {\tt (ix\_abs,iy\_abs,iz\_abs)} at this level. The software we
provide requires the 3-torus to be a cube. The side lengths input must
all be equal so that {\tt ix\_per = iy\_per = iz\_per}.  The
side-length is measured in units of octree cells at level {\tt lev} of
the octree.  The relative coordinates {\tt (ix\_rel,iy\_rel,iz\_rel)}
define the origin of a cubic subregion within the 3-torus. 
The input {\tt wn\_level\_base} is the octree level given in the
descriptor, {\tt check\_int} is the check digit in the descriptor and
{\tt phase\_name} is the name from the descriptor.  The routine does a
consistency check by first reconstituting the descriptor and then
calling the {\tt validate\_descriptor} routine described below. If it
finds a descrepancy then it simply crashes the code as something is
seriously wrong with the inputs and it is not safe to proceed.

\section{ The Random number interface. \label{mrg_app}}
Although we use the \MRG\ generator for \WNF, the routines we use are
general and can be used in principle as an interface for alternate
generators. We will describe these routines in their full generality
and only occasionally make reference to features that are particular
to our implementation.

All random number generators have the same basic structure.
Any \PRNG\ has a finite number of internal states. 
These are encoded in a {\em State-type} $\Statetype$. 
Each time the \PRNG\ is called an {\em Update transformation} 
$\Update$ is applied to the state:
\[
        \Update ( \State{\Statetype_i} )  \Rightarrow \State{ \Statetype_{i+1}}
\]
The next number in the {\em output sequence} is then generated by
applying an {\em output function } $\Output$ to the new state:
\[
       \Output ( \State{ \Statetype} ) 
        \Rightarrow
       \Real{ \Realtype }
\]
Eventually the \PRNG\ will return to one of its previous states and
the output sequence will start to repeat itself. The length of this
cycle is called the {\em period}.  The common convention is to have
the \PRNG\ generate uniformly distributed real numbers in the range
$0-1$ $\Realtype$.  More complex number distributions can be derived
from these.

\begin{table}
{
\begin{tabular}{|l|l|}
\hline
Label   & Pseudorandom generator routines  \\             
\hline
B1   &{\tt Rand\_seed}        \\     
B2   &{\tt Rand\_read}        \\            
B3   &{\tt Rand\_save}        \\ 
B4   &{\tt Rand\_load}        \\ 
B5   &{\tt Rand\_step}        \\ 
B6   &{\tt Rand\_boost}       \\
B7   &{\tt Rand\_set\_offset} \\  
B8   &{\tt Rand\_add\_offset} \\         
B9   &{\tt Rand\_mul\_offset} \\         
\hline
\end{tabular}
}
\caption{Functions and subroutines provided in the file
{\tt generic\_lecuyer.f90}.}
\label{mrg_routines}
\end{table}

In addition to being able to generate random numbers the package has
to provide some mechanism to initialise the state of the generator.
For example the user could provide a single integer that is used to
{\em seed} the generator.
\[
        \Seed ( \Integer{ \Integertype} )
        \Rightarrow
        \State{ \Statetype}
\]
The function $\Seed$ maps all valid states of the integer type
$\Integertype$ to distinct states.

 We have to define some mechanism that allows checkpointing of the
\PRNG. For example by providing mapping functions between the
state-type and integer variables.
It is quite possible that the state-type may have a greater number of
valid states than the integer type so an array of integers will required.
\[
      \Save ( \State{ \Statetype })
        \Rightarrow
      \Integer{  \Integertype ( \N )}
\]
\[
        \Load ( \Integer{ \Integertype ( \N )}  )
        \Rightarrow
        \State{ \Statetype}
\]
A \PRNG\ that implements stepping, such as \MRG, provides some
mechanism to efficiently advance the generator by an arbitrary number
of steps in the sequence.
\begin{eqnarray*}
        \Step ( \Integer{x} , \State{ \Statetype_i} ) & \Rightarrow & 
        \State{ \Statetype_{i+x} }
\end{eqnarray*}
The {\em stepping transformation} $\Step$ is parameterised by an
integer $x$ and is equivalent to the update transformation applied $x$ times.

\begin{table*}
{
\begin{tabular}{|l||l|} \hline
                & Fortran-90 \\ \hline
The State-type  & {\tt TYPE(\RNG\_state)} \\ 
The Offset-type & {\tt TYPE(\RNG\_offset)} \\
Number of integers to hold state      &
{\tt INTEGER, PARAMETER::Nstate } \\ \hline
Is stepping available      &
{\tt LOGICAL, PARAMETER::Can\_step } \\ \hline
{\tt LOGICAL, PARAMETER::Can\_reverse } \\ \hline
\end{tabular}
}\label{tab_defs}
\end{table*}

As the period of the cycle may be much greater than can be represented
by an integer an {\em offset-type} $\Offsettype$ is required to be 
able to specify all possible step lengths. This gives us a second
stepping transformation, the {\em offset transformation} $\Boost$
parameterised by an offset-type.
\[
        \Boost ( \Offset{x} ,\State{ \Statetype_i}) \Rightarrow 
   \State{\Statetype_{i+x}}
\]
Any arbitrary offset can be constructed from the following operations
\begin{enumerate}
\item  Create offset-type from integer
        \[
                \Integer{x} \Rightarrow \Offset{ \Offsettype_x}
        \]
\item Add two offsets
        \[
                \Offset{\Offsettype_x} \Add \Offset{\Offsettype_y}  
                 \Rightarrow \Offset{\Offsettype_{x+y}}
        \]
\item Multiply offset by integer
        \[
                \Offset{\Offsettype_x} \Mul \Integer{y}  
                \Rightarrow \Offset{\Offsettype_{x \times y}}
        \]
\end{enumerate}
The use of offset types may allow some of the computation involved in
the stepping operation to be pre-calculated when constructing the
offset. This may give significant savings if the offset is re-used
multiple times. 
In this case the offset could be a matrix or other operator acting on
the state. Therefore we cannot assume that it is possible to extract
the numerical value of the offset from an offset type.

Not all algorithms can be stepped efficiently though it is always
possible to iterate the generator the required number of times.

The RNG interface defines a number of constants and type definitions
that are required by the application code. These definitions must be
included as follows:
\begin{itemize}
\item      {\tt USE module-name}
\end{itemize}
Only the {\em module-name} is derived from the algorithm all other
identifiers are standardised. This requires the minimum changes to the
source code to change the algorithm used but still allows a single
program to use different generators. 

Fortran-90 allows a {\tt USE} statement to specify a rename-list so
one conforming module can be used to construct another (for example by
applying a shuffle).

The definitions provided by the package are shown in Table~\ref{tab_defs}.

The {\tt Can\_step} parameter is only {\tt .TRUE. } if the
implementation can step the generator more efficiently than just
iterating the generator. If it is {\tt .FALSE. } the stepping
functions are not available and the application must either abort or
iterate the generator itself.

The {\tt Can\_reverse} parameter is only {\tt .TRUE.} if negative
integers can be specified as input to {\tt Rand\_step} and 
{\tt Rand\_set\_offset}. If it is {\tt .FALSE. } negative inputs produce
undefined behavior. Negative offsets are needed to accurately step
``shuffled generators'' as the underlying generator must be run
backwards to initialise the shuffle-table after each stepping operation.

The routines defined in the package are as follows:
\subsection{Seeding the generator}
      \begin{proto}
        SUBROUTINE  Rand\_seed(state,n) \\
        TYPE(Rand\_state), INTENT(OUT)::state \\
        INTEGER, INTENT(IN)::n
      \end{proto}
This routine overloads the assignment operator.
Where the number of possible states is larger than the number of
possible default INTEGER values this routine attempts to produce
uncorrelated states even if this results in expensive computation.
The integer {\tt n} should be a positive integer.
\subsection{Generating Random numbers.}
The following Generic interface is used. This overloads different supported
REAL kinds and also scalar/rank-1 vector. 
      \begin{proto}
        SUBROUTINE Rand\_real(target,state) \\
        REAL, INTENT(OUT)::target \\
        TYPE(RAND\_state), INTENT(INOUT)::state
       \end{proto}
Result is uniformly distributed between zero and one, but in the case
of our implementation of \MRG\ neither zero or one is generated. The
set of supported real KINDs must include default REAL but is otherwise
un-specified.
\subsection{Saving the state}
\begin{proto}
  SUBROUTINE Rand\_save(save\_vec,x)  \\
  INTEGER, INTENT(OUT)::save\_vec(Nstate) \\
  TYPE(RAND\_state), INTENT(IN)::x
\end{proto}
This routine overloads the assignment operator.
\subsection{Loading the state.}
\begin{proto}
  SUBROUTINE Rand\_load(state,input) \\
  TYPE(RAND\_state), INTENT(OUT)::state \\
  INTEGER, INTENT(IN)::input(Nstate)
\end{proto}
This routine overloads the assignment operator.
\subsection{Stepping the generator}
\begin{proto}
  FUNCTION Rand\_step(x,n) \\
  TYPE(Rand\_state) Rand\_step \\
  TYPE(RAND\_state), INTENT(IN)::x \\
  INTEGER, INTENT(IN)::n
\end{proto}
\subsection{Jumping through the sequence}
\begin{proto}
  FUNCTION Rand\_boost(x,offset) \\
  TYPE(Rand\_state) Rand\_boost \\
  TYPE(Rand\_state), INTENT(IN)::x \\
  TYPE(Rand\_offset), INTENT(IN)::offset
\end{proto}
These routines overload the addition operator.
If stepping is not available these routines abort the program with an error.
The integer {\tt n} must not be negative unless {\tt Can\_reverse}
is true.
\subsection{Setting an offset}
\begin{proto}
  SUBROUTINE Rand\_set\_offset(offset,n) \\
  TYPE(Rand\_offset), INTENT(OUT)::offset \\
  INTEGER, INTENT(IN)::n
\end{proto}
This routine overloads the assignment operator.
The integer {\tt n} must not be negative unless {\tt Can\_reverse}
is true.
\subsection{Adding two offsets}
\begin{proto}
  TYPE(Rand\_offset) FUNCTION Rand\_add\_offset(a,b) \\
  TYPE(Rand\_offset), INTENT(IN)::a,b
\end{proto}
This routine overloads the addition operator.
\subsection{Multiplying two offsets}
\begin{proto}
  TYPE(Rand\_offset) FUNCTION Rand\_mul\_offset(a,n) \\
  TYPE(Rand\_offset), INTENT(IN)::a \\
  INTEGER, INTENT(IN)::n
\end{proto}
This routine overloads the multiplication operator.
The integer {\tt n} must not be negative.
If stepping is not available these routines abort the program with an error.

\end{document}